\documentclass[11pt,a4paper]{article}
\usepackage{jheppub}

\usepackage{amsmath}
\usepackage{amsfonts}
\usepackage{amssymb}
\usepackage{physics}

\usepackage{graphicx}
\graphicspath{{figures/}}
\usepackage{caption}
\usepackage{subcaption}

\usepackage{xcolor}
\usepackage{tikz}
\usetikzlibrary{positioning}
\usetikzlibrary{calc}

\allowdisplaybreaks

\title{Improving the performance of weak supervision searches using data augmentation}

\author[a]{Zong-En Chen,}
\author[a,b]{Cheng-Wei Chiang,}
\author[a]{and Feng-Yang Hsieh}

\affiliation[a]{Department of Physics and Center for Theoretical Physics, National Taiwan University, \\ Taipei 10617, Taiwan}
\affiliation[b]{Physics Division, National Center for Theoretical Sciences,\\ Taipei 10617, Taiwan}

\emailAdd{r10222045@ntu.edu.tw, chengwei@phys.ntu.edu.tw, f10222035@ntu.edu.tw}

\abstract{Weak supervision combines the advantages of training on real data with the ability to exploit signal properties. However, training a neural network using weak supervision often requires an excessive amount of signal data, which severely limits its practical applicability. In this study, we propose addressing this limitation through data augmentation, increasing the training data's size and diversity. Specifically, we focus on physics-inspired data augmentation methods, such as $p_{\text{T}}$ smearing and jet rotation. Our results demonstrate that data augmentation can significantly enhance the performance of weak supervision, enabling neural networks to learn efficiently from substantially less data.}

\begin{document}

\maketitle

\section{Introduction}% (fold)
\label{sec:introduction}

    In recent years, advances in machine learning have created many opportunities in collider physics. Among these advances, neural networks (NNs) have emerged as powerful tools for their exceptional performance in classification tasks. This strength naturally suggests the potential to leverage neural networks for isolating signal events from background noise in collider experiments. To train such a neural network, there are three common strategies depending on how the training data are labeled:

    \begin{enumerate}
        \item Fully supervised learning: all data are labeled.
        \item Unsupervised learning: none of the data is labeled.
        \item Weakly supervised learning: the data are labeled imperfectly.
    \end{enumerate}

    Fully supervised learning has the advantage of allowing neural networks to effectively learn distinctive signal properties from the data. However, it requires all the data to be labeled. Training data must be obtained through simulations, which can potentially contain artifacts. Unsupervised learning directly trains on real data without relying on simulations. A common approach is to use autoencoders trained with presumably mostly backgrounds. After training, the autoencoders use the reconstruction error as a test statistic to distinguish the signal from the background. However, since autoencoders train with predominantly background events, they cannot learn the distinctive signal properties. Hence, such neural networks may sometimes be poor discriminators when distinguishing signals from backgrounds~\cite{Batson:2021agz,Farina:2018fyg}.

    In contrast, weakly supervised learning can combine the advantages of both fully supervised learning (exploiting signal properties) and unsupervised learning (data-driven training). Specifically, this strategy allows the neural network to learn the signal properties for data and learn directly from real data. The weakly supervised learning approach has been applied in the experimental searches by ATLAS and CMS~\cite{ATLAS:2020iwa,CMS-PAS-EXO-22-026}.

    This study focuses on a weakly supervised learning technique called Classification Without Labels (CWoLa)~\cite{Metodiev:2017vrx}. CWoLa trains a signal/background classifier through mixed datasets. According to the Neyman-Pearson lemma~\cite{Neyman:1933wgr}, it can be shown that the optimal classifier for distinguishing between mixed datasets is as effective as the optimal classifier for distinguishing signals from backgrounds~\cite{Metodiev:2017vrx}. Our approach uses kinematic variables to define the signal and sideband regions and prepare two mixed datasets with different signal-to-background ratios for training. This setup is inspired by the CWoLa Hunting method~\cite{Collins:2018epr}.

    Even though weakly supervised learning can combine the advantages of both fully supervised and unsupervised learning, it faces practical challenges when the number of signals is limited or below a certain threshold~\cite{Collins:2021nxn,Finke:2023ltw,Freytsis:2023cjr,Beauchesne:2023vie,Cheng:2024yig,Li:2024htp}. In such cases, the neural network is unable to learn the difference between signals and backgrounds, resulting in indiscriminately cutting on both. We describe the minimum amount of signal events for the successfully trained neural network to perform better than the traditional method as the learning threshold. Unfortunately, this threshold can be greater than what would be necessary for discovery without using neural networks, thereby diminishing the practical value of the model.

    A fundamental challenge of weak supervision is that neural networks usually require a large amount of data to efficiently learn a task. To overcome this challenge, one should create a neural network that requires less real data for training or directly increase the training sample size. References~\cite{Finke:2023ltw, Freytsis:2023cjr} employed the boosted decision tree (BDT) algorithm to reduce the amount of data required. Another feasible solution~\cite{Beauchesne:2023vie,Cheng:2024yig,Li:2024htp} is pre-training a neural network using simulation data from various similar scenarios, followed by fine-tuning it with real data. This approach allows the neural network to first acquire useful knowledge from multiple scenarios and then leverage that to train on real data, making the learning process more efficient and significantly lowering the learning threshold.

    However, a potential issue with the pre-training approach is its reliance on the physics model data used for pre-training, which are based on our physical priors. Even though the data can come from a large dataset, it is still finite. The pre-training data may deviate from the real data. If the correlation between the pre-training data and the real data is not sufficiently strong, improvement in the performance may be very limited.

    To address these issues, we propose to employ data augmentation techniques~\cite{Chen:2020uds,Dolan:2021pml,Bradshaw:2019ipy,Fujimoto:2021zas,Moskowitz:2024deh}. Such techniques increase the size and diversity of the training dataset through various transformations applied to the existing ones. These transformations are inspired by human understanding of the data. By increasing the dataset size and diversity, data augmentation directly addresses the issue of limited samples. Additionally, because the augmented data are derived from the original true dataset, it is a data-driven approach that avoids the artifacts typically inherent in simulation-based approaches. Data augmentation not only increases the size of the training set but also exposes neural networks to a broader range of realistic variations without introducing synthetic artifacts. In this work, we focus on physics-inspired augmentation methods: $p_{\text{T}}$ smearing and jet rotation.

    We will demonstrate that data augmentation can reduce learning thresholds to at least half of their original value, making the neural network more sensitive to signals. Also, by combining both data augmentation methods, the neural network outperforms using individual methods alone. Besides, we present the behavior of neural networks across various augmented sample sizes, studying their asymptotic behaviors.

    This paper is organized as follows. In section~\ref{sec:hidden_valley_model}, we briefly review the Hidden Valley model as the benchmark in our later analysis. Event generation through Monte Carlo simulations, event selection criteria, datasets used in neural network training, and jet image preparation are discussed in section~\ref{sec:sample_preparation}. Section~\ref{sec:weakly_supervised_learning_with_cwola} demonstrates the original CWoLa results.  We point out a sculpting effect and discuss how to remove it. Details of data augmentation methods, the corresponding improved results, and impacts of systematic uncertainty are provided in section~\ref{sec:data_augmentation}. Finally, section~\ref{sec:conclusions} concludes our findings in this work.

% section introduction (end)

\section{Hidden Valley model}% (fold)
\label{sec:hidden_valley_model}

    To illustrate the effect of data augmentation, we take the Hidden Valley model~\cite{Carloni:2011kk,Carloni:2010tw,Beauchesne:2018myj,Albouy:2022cin} as an explicit benchmark. In this section, we briefly review this model and the model parameters considered.

    In the Hidden Valley model, a set of dark fermions is introduced, which are charged under a confining $\mathrm{SU}(3)_{\text{dark}}$ group with a confinement scale $\Lambda_\text{D}$, while remaining neutral under the Standard Model (SM) interactions. For simplicity, we assume these dark fermions have degenerate masses. The signal process considered here is $p p \to Z'$, where $Z'$ is a massive Abelian gauge boson mediating interactions between the SM and dark sectors. We assume $Z'$ has a mass of 5.5~TeV and a decay width of 10~GeV~\cite{Beauchesne:2023vie}.

    Once produced, the $Z'$ boson decays into a pair of dark quarks, $q_\text{D} \overline{q}_\text{D}$. The dark quark pairs then undergo parton showering and hadronization in the dark sector and result in collimated jets of dark hadrons, a process called ``dark showering.'' These dark hadrons include dark pseudo-scalar mesons (such as dark pions $\pi_\text{D}$) and dark vector mesons (such as dark rho mesons $\rho_\text{D}$), which decay back into SM particles through $Z'$, thereby potentially mimicking SM QCD jets in the detector. Consequently, the expected signal is a pair of jets with an invariant mass close to the mass of the $Z'$ boson.

    Following the mass relations recommended in reference~\cite{Albouy:2022cin}, we set the dark pion mass $m_{\pi_\text{D}}$ and rho meson mass $m_{\rho_\text{D}}$ as follows:
    \begin{align}
        \label{Eq:relation_of_mass}
        \frac{m_{\pi_\text{D}}}{\Lambda_\text{D}} = 5.5\sqrt{\frac{m_{q_\text{D}}}{\Lambda_\text{D}}},
        \quad
        \frac{m_{\rho_\text{D}}}{\Lambda_\text{D}} = \sqrt{5.76 + 1.5\frac{m_{\pi_\text{D}}^2}{\Lambda_\text{D}^2}},
        \quad
        m_{q_{\rm const}} = m_{q_\text{D}} + \Lambda_\text{D},
    \end{align}
    where $\Lambda_\text{D}$ is the dark confining scale, and $m_{q_{\rm const}}$ and $m_{q_\text{D}}$ are the constituent and current masses of the dark quarks, respectively.

    As in reference~\cite{Beauchesne:2023vie}, we consider two scenarios to study the effect of data augmentation. In both scenarios, $\Lambda_\text{D}$ is set to 10~GeV, and other Hidden Valley module parameters in \verb|Pythia| are set to be the same as in table~1(a) of reference~\cite{Beauchesne:2023vie}. In the first scenario, the ratio $m_{\pi_\text{D}}/\Lambda_\text{D}$ is set to 1. Here, the mass of the dark rho mesons exceeds twice the mass of the dark pions, allowing the decay process $\rho_\text{D} \to \pi_\text{D} \pi_\text{D}$, which is assumed to dominate. For simplicity, we set the branching ratio of this decay to 1, with all dark pions subsequently decaying into SM $d \overline{d}$ pairs. This is referred to as the indirect decay (ID) scenario. In the second scenario, $m_{\pi_\text{D}}/\Lambda_\text{D}$ is set to 1.8 so that the mass of the dark rho mesons is less than twice the mass of the dark pions. As a result, the decay $\rho_\text{D} \to \pi_\text{D} \pi_\text{D}$ is kinematically forbidden. For simplicity, all dark pions and dark rho mesons directly decay into SM $d \overline{d}$ pairs. This is referred to as the direct decay (DD) scenario.

% section hidden_valley_model (end)

\section{Sample preparation}% (fold)
\label{sec:sample_preparation}

    In this section, we describe the preparation of training and testing samples. Signal events are generated from a Hidden Valley model process, while the main background consists of QCD di-jet events. After selecting events based on kinematics, we prepare two mixed datasets for CWoLa training. We construct jet images as the inputs of the neural network.

    \subsection{Monte Carlo samples}% (fold)
    \label{sub:monte_carlo_samples}

        The signal process considered in this work is $pp \to Z' \to q_\text{D} \overline{q}_\text{D}$ at the CERN LHC. Signal samples are generated at leading order and hadronized by \verb|Pythia 8.307|~\cite{Sjostrand:2014zea} with the Hidden Valley model module~\cite{Carloni:2011kk,Carloni:2010tw}. The parton distribution function (PDF) set used is \verb|NN23LO1| PDF set~\cite{Ball:2012cx}. The main background is the SM QCD di-jet events $pp \to jj$. We use \verb|MadGraph5_aMC@NLO 2.7.3|~\cite{Alwall:2014hca} with the \verb|NN23LO1| PDF set~\cite{Ball:2012cx} to generate leading order samples at the parton level, followed by parton showering and hadronization with \verb|Pythia 8.307|~\cite{Sjostrand:2014zea}.

        For both signal and background samples, we consider the collisions with the center-of-mass energy 13~TeV and the luminosity $\mathcal{L} = \text{139 fb}^{-1}$, and use \verb|Delphes 3.4.2|~\cite{deFavereau:2013fsa} with the CMS default card for detector simulation. The jets are reconstructed with \verb|FastJet 3.3.2|~\cite{Cacciari:2011ma} using the anti-$k_t$~\cite{Cacciari:2008gp} algorithm with radius $R = 0.8$. This larger radius is used to accommodate the signal jets from dark showering. According to our simulations, this ensures that at least $90\%$ of the jet constituents are included within the radius. Only jets with a transverse momentum of $p_\text{T}\ge \text{20 GeV}$ are considered.

        After the detector simulation, we focus on the events that contain at least two jets. Each of the two leading jets needs to have $p_{\text{T}} > \text{750 GeV}$ and be within the range $\abs{\eta} < 2$. We define the signal and sideband regions based on the invariant mass $m_{jj}$ of the two leading jets.
        \begin{itemize}
            \item Signal Region (SR): It contains events with $m_{jj} \in [4700, 5500] \text{ GeV}$.
            \item Sideband Region (SB): It contains events with $m_{jj} \in [4400, 4700] \cup [5500, 5800] \text{ GeV}$.
        \end{itemize}

    % subsection monte_carlo_samples (end)
    \subsection{Datasets}% (fold)
    \label{sub:datasets}

        We prepare two mixed datasets for CWoLa training. These two mixed datasets come from the experimental data in the SR and SB. In our study, we utilize simulated samples and manually mix the signal and background events for the neural network training. With the assumed luminosity $\mathcal{L} = \text{139 fb}^{-1}$, the cross-sections of the main background process in the SR and SB are about $\text{136.1 fb}$ and $\text{145.6 fb}$, respectively. Given this, the number of events in the SR and SB are about 19k and 20k, respectively. We then vary the number of signal events in the mixed datasets to observe how the performance of neural networks is affected by the signal-to-background ratio in the training data. Moreover, this enables us to determine the learning thresholds of the neural network.

        Among the prepared training samples, 80\% of the dataset is used for training and 20\% for validation. For the testing part, we prepare a pure dataset to evaluate the model's performance, which consists of 20k signal events and 20k background events in the signal region.

        To evaluate the robustness of the neural networks, we prepare larger datasets for training and testing. The background dataset contains 200k events, while the signal datasets for ID and DD include 60k events each. All these events pass the selection criteria. We re-sample the events from these larger datasets to prepare distinct samples for training and testing.

    % subsection datasets (end)
    \subsection{Jet images}% (fold)
    \label{sub:jet_images}

        The inputs of neural networks are jet images~\cite{Kasieczka:2019dbj,deOliveira:2015xxd, Kasieczka2017nv}. We construct jet images from the event passing the kinematic requirements described in section~\ref{sub:monte_carlo_samples}. The jet image is constructed for each jet separately so that we can obtain two for each event. The following preprocessing steps are applied to jet constituents to construct the jet image:
        \begin{enumerate}
            \item Translation: Compute the $p_{\text{T}}$-weighted center in the $\left( \eta,\phi \right) $ coordinates, then shift this point to the origin.
            \item Orientation: Rotate the highest intensity axis to align with the $\eta$ axis.
            \item Flipping: Flip the highest $p_{\text{T}}$ constituent particle to the first quadrant.
            \item Pixelation: Pixelate in a $\eta \in [-1,1],\ \phi \in [-1,1]$ box, with $25 \times 25$ pixels \footnote{Additionally, we have tried the resolution of $75\times75$ pixels with data augmentation methods and observed similar performance improvement in the neural networks. To avoid unnecessary duplication, we present exclusively the results obtained by using the resolution of $25\times25$ pixels.}.
        \end{enumerate}

        In this work, we utilize the $m_{jj}$ variable to construct two mixed datasets. This could inadvertently lead to a sculpting effect, which refers to the phenomenon that the neural network does not learn properly the differences between the signal and background samples but learns the distinction in the definitions of the SR and SB. In our case, this implies that the classifier could learn the di-jet invariant mass information from the input samples and use it as a discriminator. However, an essential assumption of the CWoLa Hunting method is that the input data distributions should be the same in the SR and SB, except for the variable used to define the two regions. Therefore, the inputs utilized by the classifier should be independent of $m_{jj}$.

        To remove the dependence of the input samples on $m_{jj}$, we utilize normalization techniques that standardize the jet images to remove the difference in input data distributions between the SR and SB. We calculate the mean and standard deviation of the jet image transverse momentum and use these values to standardize each jet image using two normalization schemes:
        \begin{enumerate}
            \item Jet Normalization (JN): Each jet image is standardized individually. This method removes the $p_{\text{T}}$ differences between the leading and sub-leading jets.

            \item Event Normalization (EN): We compute the mean and standard deviation of both leading jet images, then standardize the two jet images using these values. This method removes the difference among various events and also keeps the difference between the leading and sub-leading jets.
        \end{enumerate}

    % subsection jet_images (end)
% section sample_preparation (end)

\section{Weakly supervised learning with CWoLa}% (fold)
\label{sec:weakly_supervised_learning_with_cwola}

    In this section, we demonstrate the details of our CWoLa training setup, examine the sculpting effect, and present the results of the NN selection.

    \subsection{Model structure and training setup}% (fold)
    \label{sub:model_structure_and_training_setup}

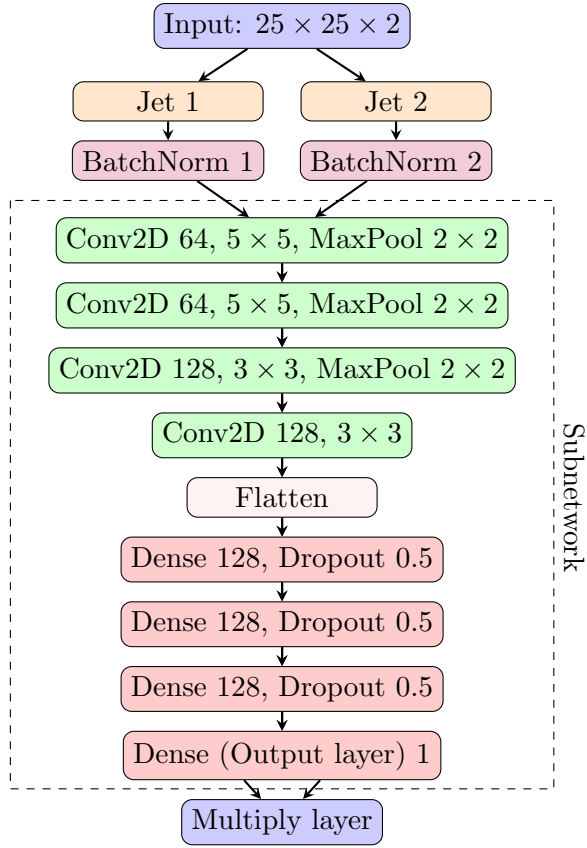
\begin{figure}[!ht]
        \centering
        \begin{tikzpicture}[
            layer/.style={draw, rounded corners, minimum width=2.5cm, text centered},
            arrow/.style={-stealth, thick},
            node distance=0.25cm,
        ]
        % Input layer
        \node (input) [layer, fill=blue!20,] {Input: $25 \times 25 \times 2$};

        % Jet 1 layers
        \node (jet1) [layer, fill=orange!20] at ($(input)+(-1.5,-1)$) {Jet 1};
        \node (bn1) [below=of jet1, layer, fill=purple!20] {BatchNorm 1};

        % Jet 2 layers
        \node (jet2) [layer, fill=orange!20] at ($(input)+(1.5,-1)$) {Jet 2};
        \node (bn2) [below=of jet2, layer, fill=purple!20] {BatchNorm 2};

        % Conv1_1 layer positioned between bn1 and bn2
        \node (conv1_1) [below=0.75cm of $(bn1)!0.5!(bn2)$, layer, fill=green!20] {Conv2D 64, $5 \times 5$, MaxPool $2 \times 2$};

        % Other Conv layers
        \node (conv2_1) [below=of conv1_1, layer, fill=green!20] {Conv2D 64, $5 \times 5$, MaxPool $2 \times 2$};
        \node (conv3_1) [below=of conv2_1, layer, fill=green!20] {Conv2D 128, $3 \times 3$, MaxPool $2 \times 2$};
        \node (conv4_1) [below=of conv3_1, layer, fill=green!20] {Conv2D 128, $3 \times 3$};

        % Dense layers
        \node (flatten1) [below=of conv4_1, layer, fill=pink!20] {Flatten};
        \node (dense1_1) [below=of flatten1, layer, fill=red!20] {Dense 128, Dropout 0.5};
        \node (dense2_1) [below=of dense1_1, layer, fill=red!20] {Dense 128, Dropout 0.5};
        \node (dense3_1) [below=of dense2_1, layer, fill=red!20] {Dense 128, Dropout 0.5};
        \node (output1) [below=of dense3_1, layer, fill=red!20] {Dense (Output layer) 1};

        % Multiply layer
        \node (multiply) [below=of output1, layer, fill=blue!20] {Multiply layer};

        \coordinate (p1) at ($(bn1)!0.5!(conv1_1)$);
        \coordinate (p2) at ($(output1)!0.5!(multiply)$);
        \coordinate (subnet1) at ($(p1-|conv3_1.west)+(-0.5,0)$);
        \coordinate (subnet2) at ($(p2-|conv3_1.east)+(0.5,0)$);

        \node[rotate=-90, above] (subnet) at (flatten1-|subnet2) {Subnetwork};

        % Arrows from Input to Jets
        \draw [arrow] (input) --  (jet1);
        \draw [arrow] (input) --  (jet2);

        % Arrows for BN1
        \draw [arrow] (jet1) -- (bn1);
        \draw [arrow] (bn1) -- (conv1_1);
        \draw [arrow] (conv1_1) -- (conv2_1);
        \draw [arrow] (conv2_1) -- (conv3_1);
        \draw [arrow] (conv3_1) -- (conv4_1);
        \draw [arrow] (conv4_1) -- (flatten1);
        \draw [arrow] (flatten1) -- (dense1_1);
        \draw [arrow] (dense1_1) -- (dense2_1);
        \draw [arrow] (dense2_1) -- (dense3_1);
        \draw [arrow] (dense3_1) -- (output1);

        % Arrows for BN2
        \draw [arrow] (jet2) -- (bn2);
        \draw [arrow] (bn2) -- (conv1_1);
        \draw [arrow] (conv1_1) -- (conv2_1);
        \draw [arrow] (conv2_1) -- (conv3_1);
        \draw [arrow] (conv3_1) -- (conv4_1);
        \draw [arrow] (conv4_1) -- (flatten1);
        \draw [arrow] (flatten1) -- (dense1_1);
        \draw [arrow] (dense1_1) -- (dense2_1);
        \draw [arrow] (dense2_1) -- (dense3_1);
        \draw [arrow] (dense3_1) -- (output1);

        % Arrows from output1 to multiply
        \draw [arrow] (output1.south) ++(-0.5,0) -- (multiply);
        \draw [arrow] (output1.south) ++(0.5,0) -- (multiply);

        \draw [dashed] (subnet1) rectangle (subnet2);
        \end{tikzpicture}
        \caption{The architecture of the neural network and model hyperparameters.}
        \label{fig:nn_architecture}
        \end{figure}

        Figure~\ref{fig:nn_architecture} shows the architecture of the neural network used in our study. The neural network is implemented in \verb|Keras|~\cite{chollet2015keras} with the \verb|TensorFlow|~\cite{Abadi:2016kic} backend. The network takes two jet images as input for an event. Jet images are fed to the batch normalization layers first and then sent to the subnetwork. The subnetwork part consists of four convolutional layers, each followed by a max-pooling layer except for the last one. These convolutional layers extract features from the jet images. Four dense layers further process the features obtained from the convolutional layers and enable the network to perform the classification task. Dropout layers with a dropping rate of 0.5 are applied to the first three dense layers to prevent overfitting. The ReLU activation function is used in all convolutional layers and the first three dense layers, and the Sigmoid function is used in the last dense layer. The final output is obtained by multiplying the two jet image outputs of the subnetwork.

        The loss function is the binary cross entropy. The \verb|Adam| optimizer is used to minimize the loss value. The learning rate is $10^{-4}$, and the batch size is 512. To prevent over-training issues, we employ the early stopping technique with a patience of 10.

    % subsection model_structure_and_training_setup (end)
    \subsection{Sculpting effect}% (fold)
    \label{sub:sculpting_effect}

        To investigate the presence of the sculpting effect, we train a neural network with the datasets, which only consist of background events in the SR and SB. We use jet images, both with and without normalization techniques. The neural network assigns a score, $p_{\text{event}}$, to each event. We then apply a cut on $p_{\text{event}}$, referred to as the NN cut, which requires $p_{\text{event}}$ to exceed a threshold. Using this approach, we obtain the $m_{jj}$ distributions and the NN cut passing efficiency, $\varepsilon$, as a function of $m_{jj}$.

\begin{figure}[!ht]
            \centering
            \includegraphics[width=\textwidth]{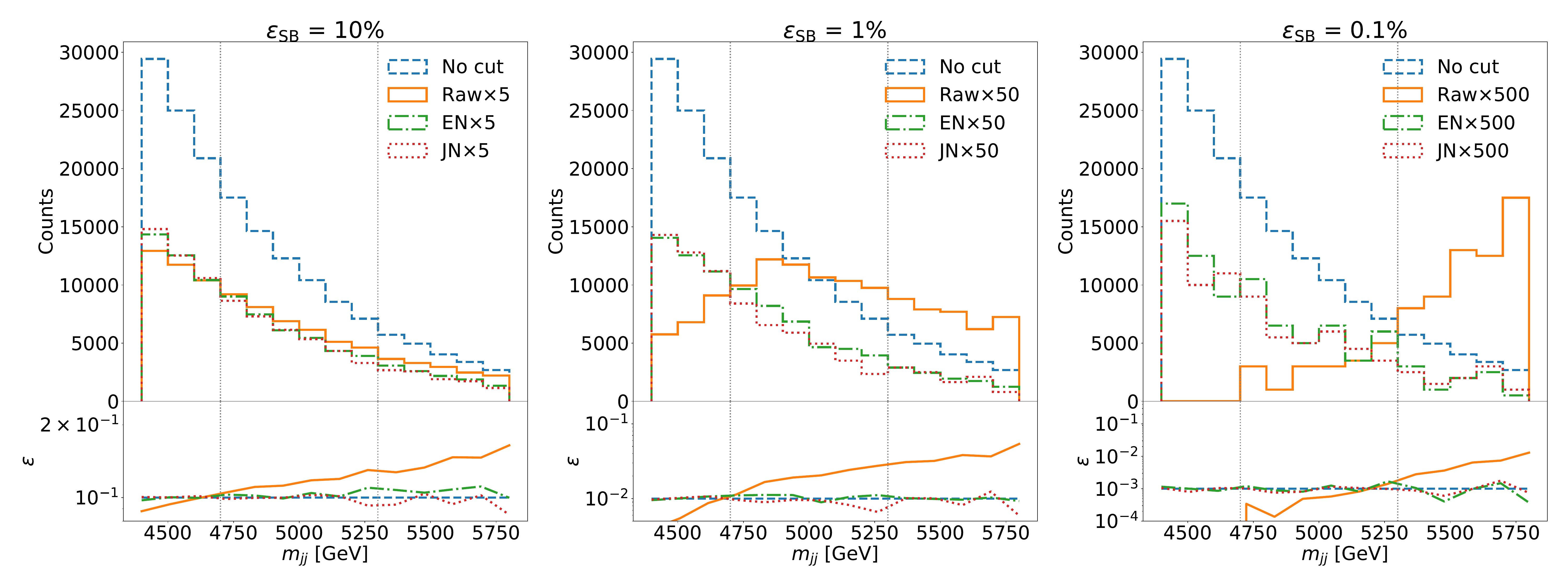}
            \caption{The invariant mass $m_{jj}$ histogram and the NN cut passing efficiency $\varepsilon$ as functions of $m_{jj}$, with different sideband efficiencies $\varepsilon_\text{SB}$.}
            \label{fig:mass-cut-ZN-2-method}
        \end{figure}

        The $m_{jj}$ distributions and NN cut passing efficiency are presented in figure~\ref{fig:mass-cut-ZN-2-method}. For comparison, we show results for events without applying the NN cut, labeled as ``No cut,’’ those for events applied with the NN cut without employing any normalization, labeled as ``Raw,'' and those for events using the JN and EN schemes applied with the NN cut. To see the results of applying the NN cut more clearly, those event counts are multiplied by 5, 50, and 500 respectively for the plots with the sideband efficiencies of $10\%$, $1\%$, and $0.1\%$. Our results indicate that both the JN and EN schemes effectively mitigate the sculpting effect. In contrast, without normalization, the neural network exhibits a bias toward keeping higher $m_{jj}$ events, thereby introducing the sculpting effect.

        Furthermore, when signal events are included, we observe that the EN scheme achieves better training performance than the JN scheme. This improvement arises because EN preserves the differences between the leading and sub-leading jets, enhancing the neural network’s ability to distinguish events. We therefore choose to apply the EN scheme in all our samples for subsequent analyses.

    % subsection sculpting_effect (end)
    \subsection{Results of CWoLa}% (fold)
    \label{sub:results_of_cwola}

        After training, we compute the signal efficiency $\varepsilon_\text{s}$ with a given background efficiency $\varepsilon_\text{b}$ from the receiver operating characteristic (ROC) curve using the testing data mentioned in section~\ref{sub:datasets}. The numbers of signal and background events passing the NN cut are determined respectively as $s=s_0\varepsilon_{\text{s}}$ and $b=b_0\varepsilon_{\text{b}}$, where $s_0$ and $b_0$ denote respectively the numbers of signal and background events before the NN cut. The sensitivity is then calculated as~\cite{ATLAS:2020yaz} :
        \begin{equation}\label{eq:sensitivity}
            \sigma = \sqrt{2\left((N_{\text{s}}+N_{\text{b}})\log(\frac{N_{\text{s}}}{N_{\text{b}}}+1)-N_{\text{s}} \right)},
        \end{equation}
        where $N_{\text{s}}$ and $N_{\text{b}}$ refer to the numbers of signal and background events, respectively.  Additionally, we re-sample the training and testing data, retrain the neural network 10 times, and compute the mean and standard deviation of the resulting sensitivities to examine the robustness of the neural network.

\begin{figure}[!ht]
            \centering
            \subfloat[ID]{
                \includegraphics[width=0.98\textwidth]{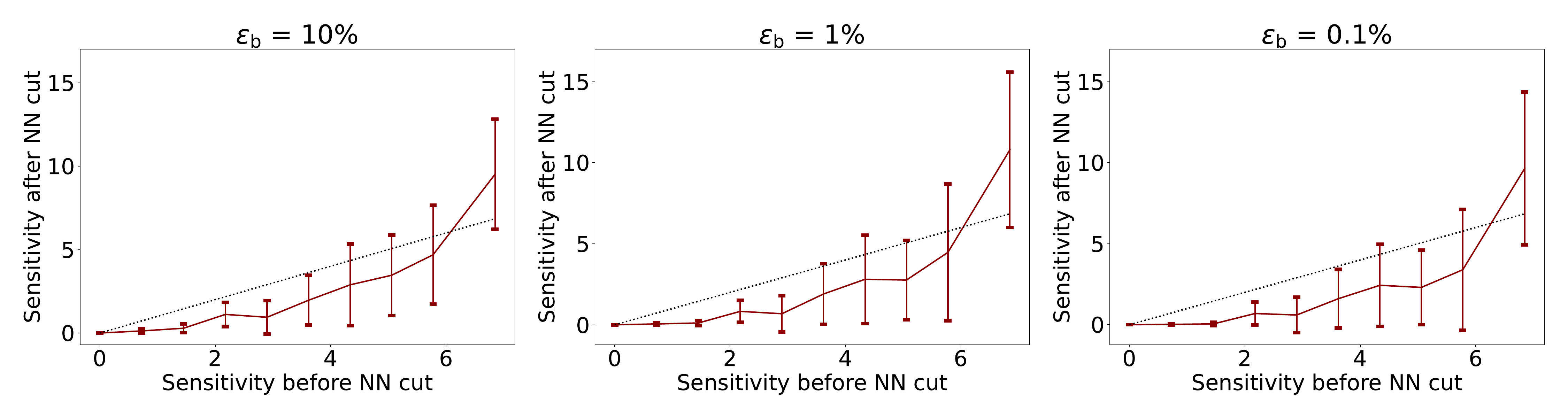}
            } \\
            \subfloat[DD]{
                \includegraphics[width=0.98\textwidth]{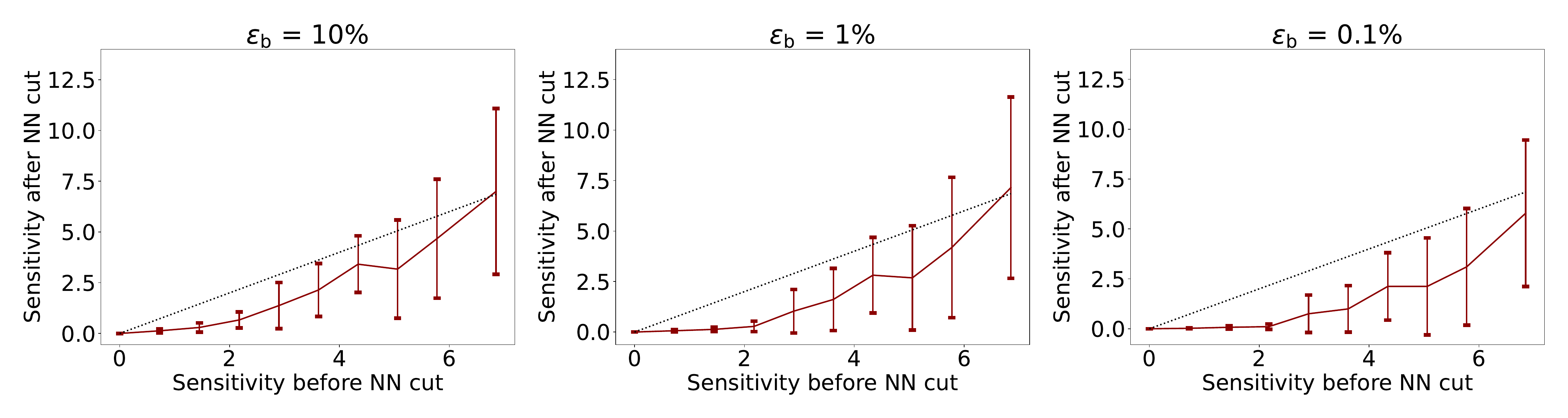}
            } \\
            \caption{The sensitivities before and after the NN selection. The gray dotted line represents the sensitivity before NN selection. The error bar is the standard deviation of 10 times training.}
            \label{fig:result_CWoLa}
        \end{figure}

        Figure~\ref{fig:result_CWoLa} shows the sensitivity improvement of the CWoLa approach for the ID and DD scenarios of our benchmark model. In both scenarios, the learning thresholds exceed $5\sigma$, diminishing the neural network's practicality. Additionally, the large standard deviation in sensitivity indicates that the neural network is largely unstable. When the amount of signals is below the learning thresholds, the neural network cannot obtain sufficient information to learn to distinguish signals from backgrounds. As a result, it indiscriminately cuts both signal and background events, yielding worse performance than if no cut is applied.

    % subsection results_of_cwola (end)
% section weakly_supervised_learning_with_cwola (end)

\section{Data augmentation}% (fold)
\label{sec:data_augmentation}

    To improve the performance of neural networks, one elementary method is using larger datasets for training. However, collecting sufficient data for training is sometimes very challenging in practice. In such cases, data augmentation becomes a powerful tool to address this limitation by increasing the size and diversity of the training data. The augmentation techniques are typically based on our understanding of the data and usually follow principles consistent with physical laws. More samples are thus obtained through specific transformations that preserve the sample label on the original dataset. These additional samples can help enhance the learning of neural networks for the task of interest.

    In the CWoLa Hunting study, the numbers of events in the SR and SB are limited by the real data under a given luminosity of the collider. Consequently, the available training data size may not be sufficiently large so that the neural network cannot effectively learn for a task. We therefore employ data augmentation techniques to overcome this challenge.

    \subsection{Data augmentation methods}% (fold)
    \label{sub:data_augmentation_methods}

        While there are numerous augmentation methods in the field of computer vision~\cite{wang2024comprehensivesurveydataaugmentation}, we focus on physics-inspired techniques related to our study. We implement three methods\footnote{Additionally, we have applied $\eta-\phi$ smearing and Gaussian noise to jet images and observed essentially no improvement.}: (i) $p_{\text{T}}$ smearing, (ii) jet rotation, and (iii) the combination of the previous two. Such methods are inspired by reference~\cite{Dillon:2023zac}, which considers the augmentations that capture the symmetries of the physical events and the experimental resolution or statistical fluctuations in the detector.

        The $p_{\text{T}}$ smearing method is used to simulate detector resolution effects on the transverse momentum of jet constituents. This method resamples the transverse momentum $p_{\text{T}}$ of jet constituents according to the normal distribution:
        \begin{equation}
            p_{\text{T}}' \sim \mathcal{N}\left( p_{\text{T}}, f(p_{\text{T}}) \right), \quad f(p_{\text{T}}) = \sqrt{0.052 p_{\text{T}}^2 + 1.502p_{\text{T}}},
        \end{equation}
        where $p_{\text{T}}'$ is the augmented transverse momentum, and $f\left( p_\text{T} \right) $ is the energy smearing function applied by \verb|Delphes| (the $p_{\text{T}}$'s are normalized in units of GeV). The preprocessing is applied after the $p_{\text{T}}$ smearing augmentation. This augmentation helps the model consider the detector effects. It has the effect of making the training results more robust.

        The jet rotation method rotates each jet with respect to its center by a random angle $\theta \in [-\pi, \pi]$ to enlarge the diversity of training datasets. More specifically, the $(\eta', \phi')$ coordinates of a jet constituent after preprocessing are rotated as follows:
        \begin{equation}
            \eta'' = \eta'\cos\theta - \phi\sin\theta, \quad \phi'' = \eta'\sin\theta + \phi'\sin\theta,
        \end{equation}
        where $(\eta'', \phi'')$ are the rotated coordinates. We allow the two leading jets in an event to be rotated by different angles, thereby further increasing the diversity of the training dataset. Note that jet rotation is applied before the pixelation step. The complete workflow for preparing jet images with this augmentation consists of the following steps: translation, orientation, flipping, jet rotation, and finally pixelation.

        We note in passing that we have tested other ranges of jet rotation angles, including $[-\pi/6, \pi/6]$, $[-\pi/3, \pi/3]$, and $[-\pi/2, \pi/2]$. Our results show that the training performance improves as the range of rotation angles increases, with the range of $[-\pi, \pi]$ yielding the best results. Therefore, we will focus exclusively on the $[-\pi, \pi]$ range in this work.

        Although applying the jet rotation after the preprocessing may seem redundant since the preprocessing aligns jet orientation and removes rotational symmetry to simplify the training as well as testing, our simulations indicate that training performance can be enhanced even without the orientation step in preprocessing as long as there is a sufficiently large dataset, thus motivating us to consider the jet rotation augmentation. Preprocessing is useful when considering small datasets, as it makes training possible by simplifying jet orientations. However, we have observed that with larger datasets, training can still be successful even without orientation preprocessing and, in most cases under consideration, removing the orientation preprocessing leads to better results. As such, we apply jet rotations to enlarge the diversity of training samples and expect that a broader range of jet configurations can improve the training performance.

        The third augmentation method that we have considered is the combination of $p_{\text{T}}$ smearing and jet rotation. They are applied sequentially. The complete workflow consists of the following steps: $p_{\text{T}}$ smearing, translation, orientation, flipping, jet rotation, and finally pixelation. This combined approach produces jet images with variations in both momentum resolution and angular position while preserving the essential jet structure.

\begin{figure}[!ht]
            \centering
            \subfloat[Original jet image]{
                \includegraphics[width=0.47\textwidth]{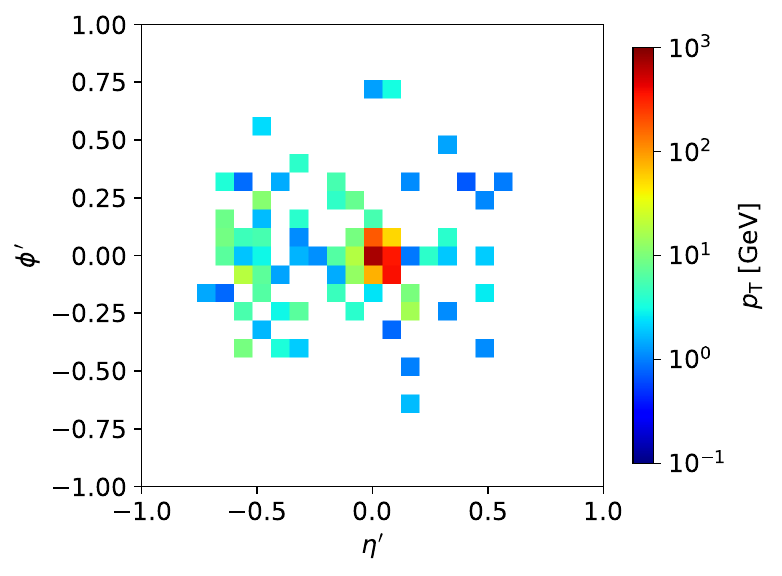}
            }
            \subfloat[$p_{\text{T}}$ smearing]{
                \includegraphics[width=0.47\textwidth]{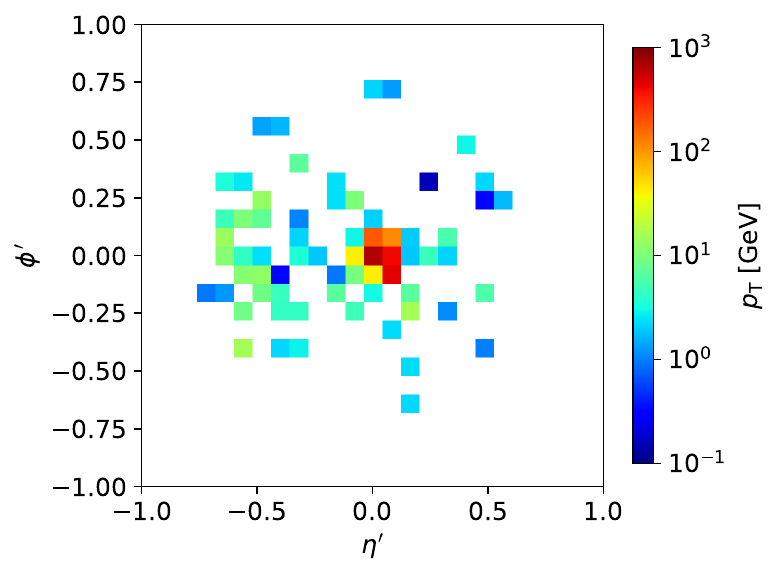}
            } \\
            \subfloat[Jet rotation]{
                \includegraphics[width=0.47\textwidth]{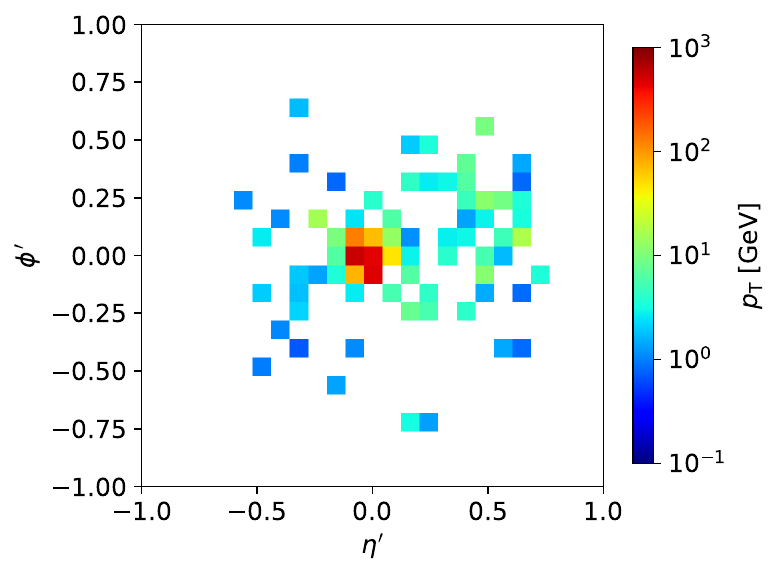}
            }
            \subfloat[$p_{\text{T}}$ smearing + jet rotation]{
                \includegraphics[width=0.47\textwidth]{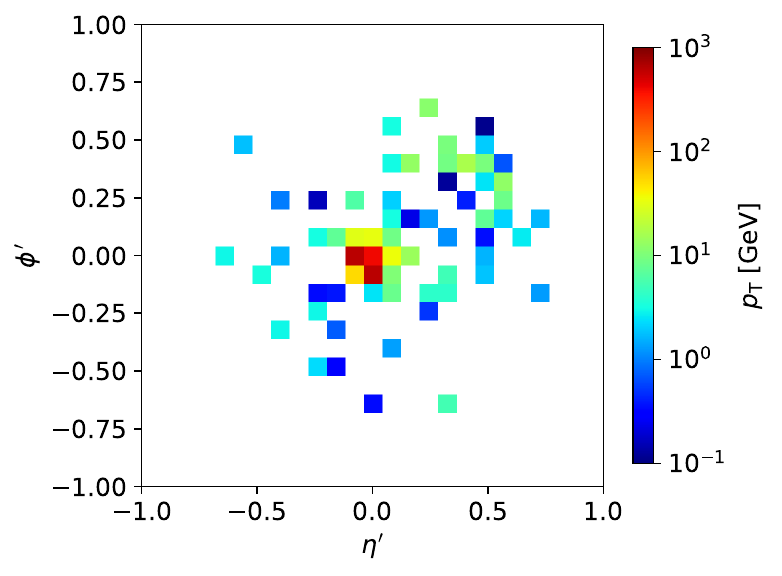}
            }
            \caption{The jet images before and after different data augmentation methods.}
            \label{fig:jet_image_before_after_augmentation}
        \end{figure}

        Figure~\ref{fig:jet_image_before_after_augmentation} shows a jet image before and after different augmentation methods. Plot (a) is the original preprocessed jet image. Plot (b) shows the jet image with $p_{\text{T}}$ smearing. Although $p_\text{T}$ smearing only modifies the transverse momentum of jet constituents, the preprocessing shifts the jet image based on $p_\text{T}$. Thus, the pixels of the image differ from the original one not only in intensity but also slightly in position. Plot (c) is the jet image after a jet rotation. Since the jet rotation only modifies the $(\eta', \phi')$ coordinates, the jet image only differs by an angle $\theta$ from plot (a) but with the same intensity. Plot (d) shows the jet image with both $p_{\text{T}}$ smearing and jet rotation. In this case, the new image has different angular position and intensity, but the overall pattern remains consistent with the original image.

    % subsection data_augmentation_methods (end)
    \subsection{Impacts of data augmentation}% (fold)
    \label{sub:impact_of_data_augmentation}

\begin{figure}[!ht]
            \centering
            \subfloat[ID, +5 augmentation]{
                \includegraphics[width=0.98\textwidth]{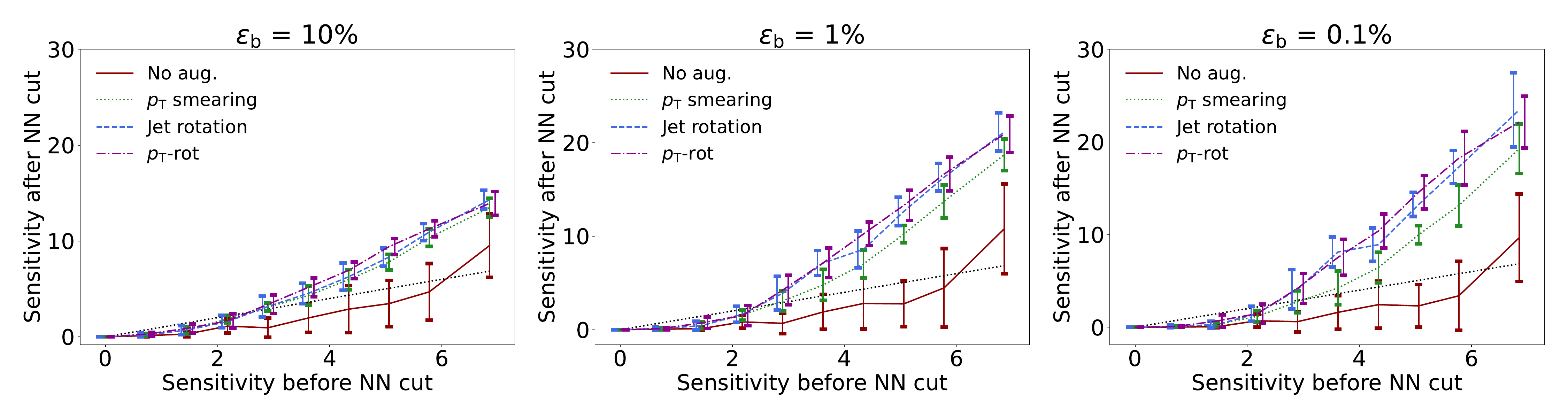}
            } \\
            \subfloat[DD, +5 augmentation]{
                \includegraphics[width=0.98\textwidth]{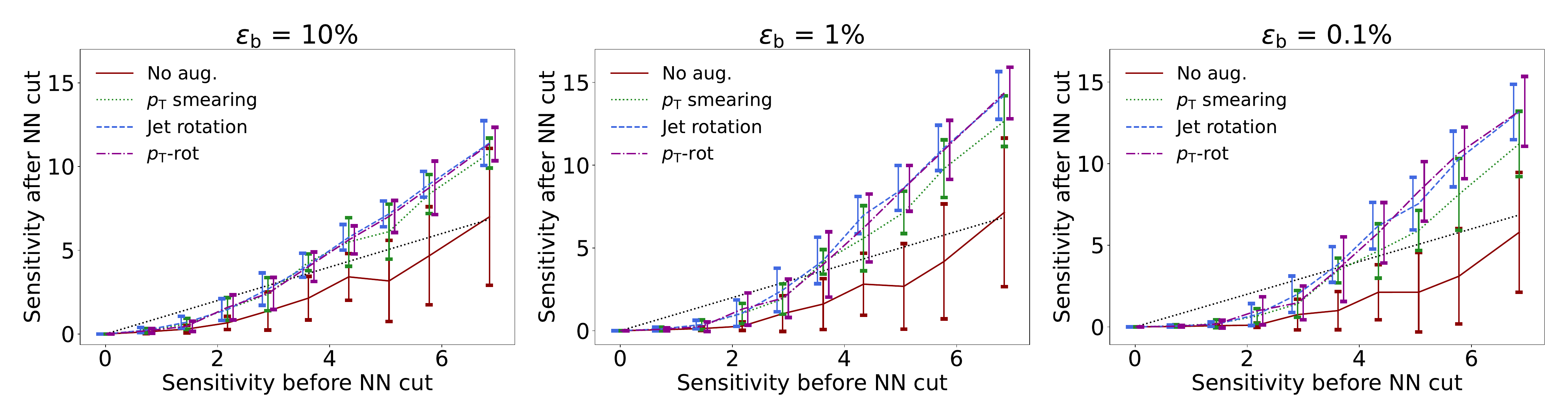}
            }
            \caption{The sensitivities before and after the NN selection. The gray dotted line represents the sensitivity before NN selection. The error bar is the standard deviation of 10 times training. The ``$p_{\text{T}}$-rot'' means the ``$p_{\text{T}}$ smearing + jet rotation'' augmentation method.}
            \label{fig:sensitivity_improvement_aug_5}
        \end{figure}

        Figure~\ref{fig:sensitivity_improvement_aug_5} shows the sensitivity improvement with different data augmentation methods for the ID and DD scenarios with different background efficiencies. Here, we consider the ``+5 augmentation,'' which means that the training dataset consists of the original data plus 5 augmented versions. As seen in the plots, even with just +5 augmentation, the model's performance significantly improves. The learning thresholds are reduced from approximately $6\sigma$ to $3\sigma$ for both scenarios and the fluctuations in the sensitivity after the data augmentation are reduced to about a half.

        Among the three augmentation methods, the ``$p_{\text{T}}$ smearing + jet rotation'' approach performs best. Because the combined approach can introduce greater diversity to the training dataset, this approach helps the neural network learn the differences between signal and background events more efficiently.

\begin{figure}[!ht]
            \centering
            \subfloat[ID]{
                \includegraphics[width=0.98\textwidth]{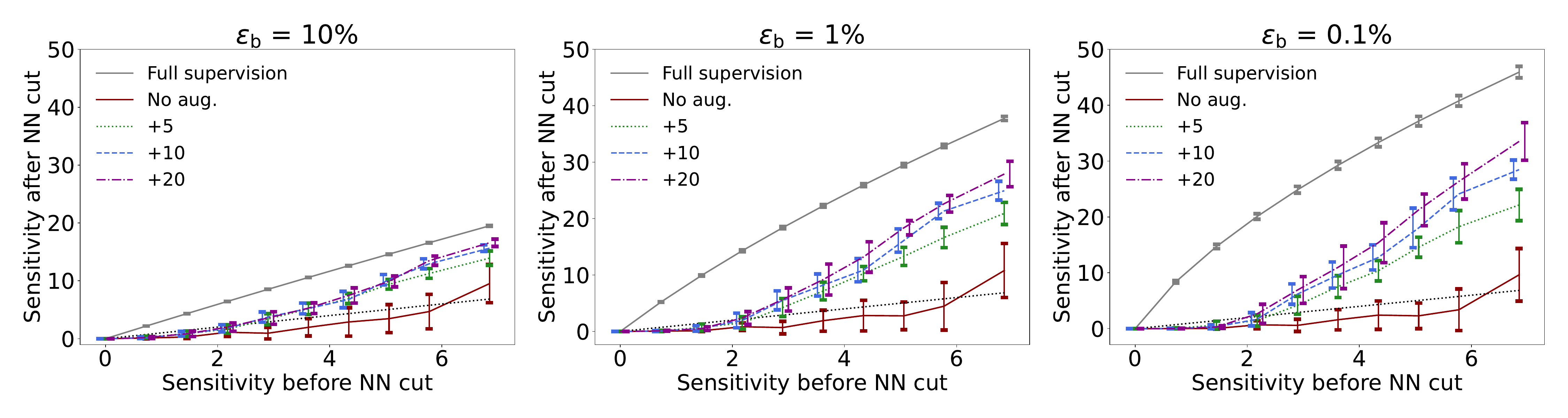}
            } \\
            \subfloat[DD]{
                \includegraphics[width=0.98\textwidth]{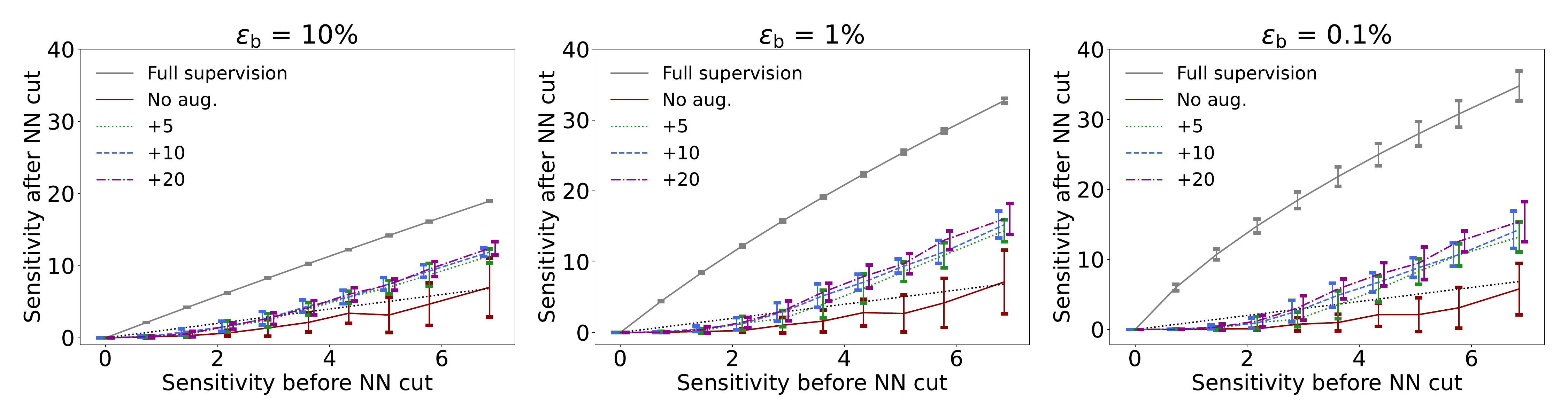}
            }
            \caption{The sensitivities before and after the NN selection with the ``$p_{\text{T}}$ smearing + jet rotation'' augmentation method. The gray dotted line represents the sensitivity before NN selection. The error bar is the standard deviation of 10 times training.}
            \label{fig:sensitivity_improvement_aug_5_10_20_fs}
        \end{figure}

        Figure~\ref{fig:sensitivity_improvement_aug_5_10_20_fs} compares the sensitivity improvements across various sizes of augmented datasets and the fully supervised case. Here, we focus on the ``$p_{\text{T}}$ smearing + jet rotation'' augmentation method. The curves for fully supervised learning represent the optimal performance of the neural network and can serve as a benchmark in distinguishing signals from backgrounds. As expected, the neural networks perform better when we increase the training sample size. However, their performance remains below that of fully supervised learning. This is because the augmented datasets are mixed, and the information on signal events is limited. Consequently, the neural network can not extract all the necessary features for optimal classification.

    % subsection impact_of_data_augmentation (end)
    \subsection{Asymptotic behavior}% (fold)
    \label{sub:asymptotic_behavior}

\begin{figure}[!ht]
            \centering
            \subfloat[ID]{
                 \includegraphics[width=0.97\textwidth]{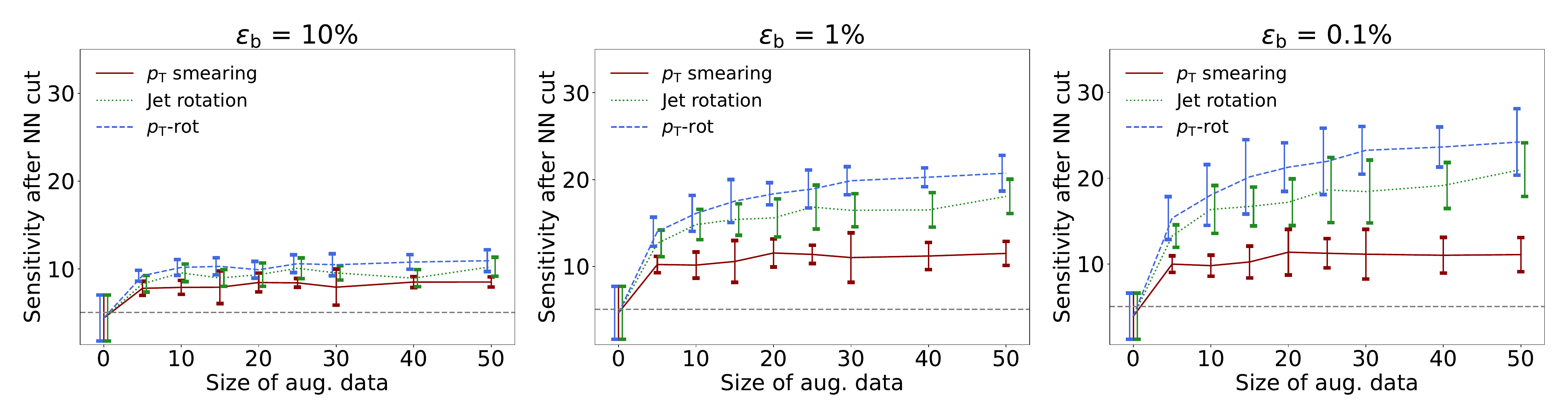}
            } \\
            \subfloat[DD]{
                 \includegraphics[width=0.97\textwidth]{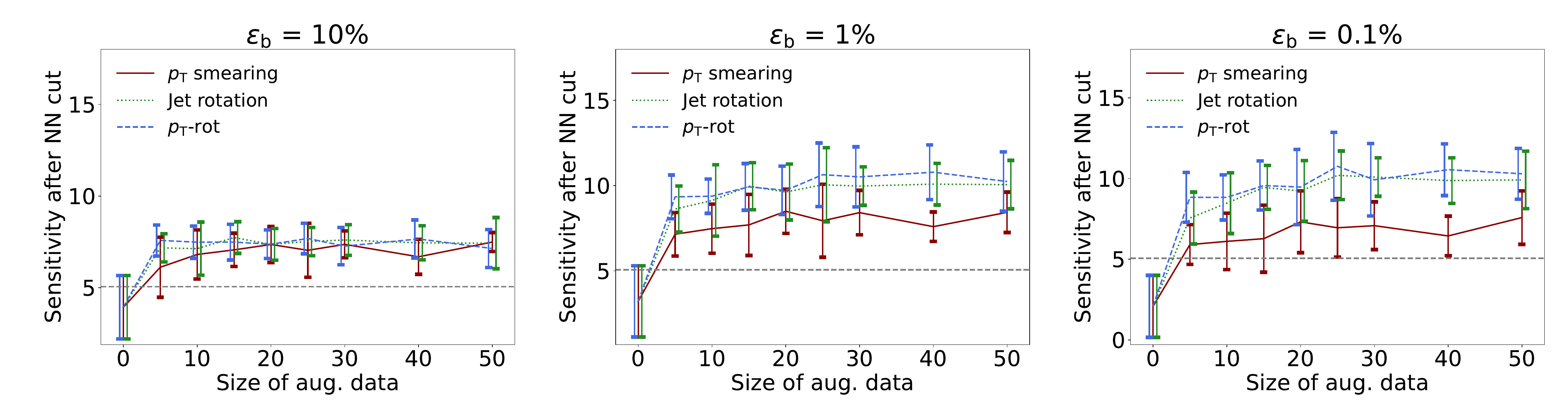}
            }
            \caption{The sensitivities after the NN selection as a function of the size of augmented data. Here we fix the sensitivity before the NN selection at $5$. The horizontal gray dashed line represents the sensitivity before the NN selection. The error bar is the standard deviation of 10 times training.}
            \label{fig:sensitivity_improvement_5_aug_50}
        \end{figure}

        To explore the behavior and limit in the performance of the neural networks with augmented datasets, we enlarge the training sample size through different data augmentation techniques.  We start with two datasets where the signal sensitivity is set to $5$ before applying the NN selection in both ID and DD scenarios.  We then augment the datasets to different sizes using the methods mentioned above. Figure~\ref{fig:sensitivity_improvement_5_aug_50} shows the sensitivity improvement with different augmented sample sizes. Again, the ``$p_{\text{T}}$ smearing + jet rotation'' method performs best among the three augmentation methods. The sensitivity improvement of the $p_{\text{T}}$ smearing method saturates the first, usually around +10 to +15 augmentation. The jet rotation and the combined methods saturate after approximately +30 augmentation in the ID scenario and even earlier in the DD scenario. Also, we have tried the original sensitivity set to be $3$. Such conclusions are the same for both cases where the original sensitivity is $3$ and $5$. This indicates that a small sample augmentation can already boost the sensitivity significantly and that there is no point in enlarging the dataset indefinitely.

    % subsection asymptotic_behavior (end)
    \subsection{Impacts of systematic uncertainty}% (fold)
    \label{sub:impacts_of_systematic_uncertainty}

        Another question that needs to be addressed is whether data augmentation can also improve the neural network's performance in the presence of systematic uncertainty. To investigate their impact, we use the following equation, modified from equation~\eqref{eq:sensitivity} to account for the effect of systematic uncertainty, to estimate the sensitivity~\cite{ATLAS:2020yaz}:
        \begin{align}
            \label{eq:sensitivity_systematic}
            \overline\sigma = \sqrt{2
            \left(  (N_{\text{s}}+N_{\text{b}})\log
                    \left[
                    \frac{(N_{\text{s}}+N_{\text{b}})(N_{\text{b}}+\sigma_{\text{b}}^2)}{N_{\text{b}}^2+(N_{\text{s}}+N_{\text{b}})\sigma_{\text{b}}^2}
                    \right]
                    -\frac{N_{\text{b}}^2}{\sigma_{\text{b}}^2}\log
                    \left[
                    1+\frac{\sigma_{\text{b}}^2 N_{\text{s}}}{N_{\text{b}}(N_{\text{b}}+\sigma_{\text{b}}^2)}
                    \right]
            \right)},
        \end{align}
        where $\sigma_{\text{b}}$ is the systematic uncertainty of the background. In the limit of $\sigma_{\text{b}}\to0$, equation~\eqref{eq:sensitivity_systematic} reduces to equation~\eqref{eq:sensitivity}. With a nonzero relative systematic uncertainty of the background ${\sigma_{\text{b}}}/{N_{\text{b}}}$, the neural network's performance becomes worse than before.

\begin{figure}[!ht]
            \centering
            \subfloat[ID, +5 augmentation]{
                \includegraphics[width=0.98\textwidth]{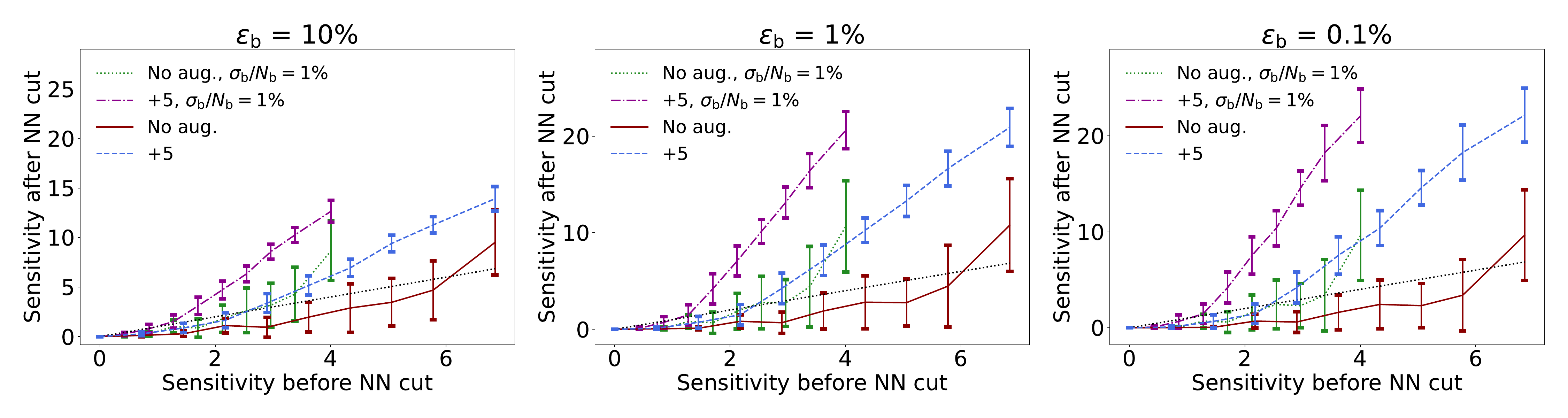}
            } \\
            \subfloat[DD, +5 augmentation]{
                \includegraphics[width=0.98\textwidth]{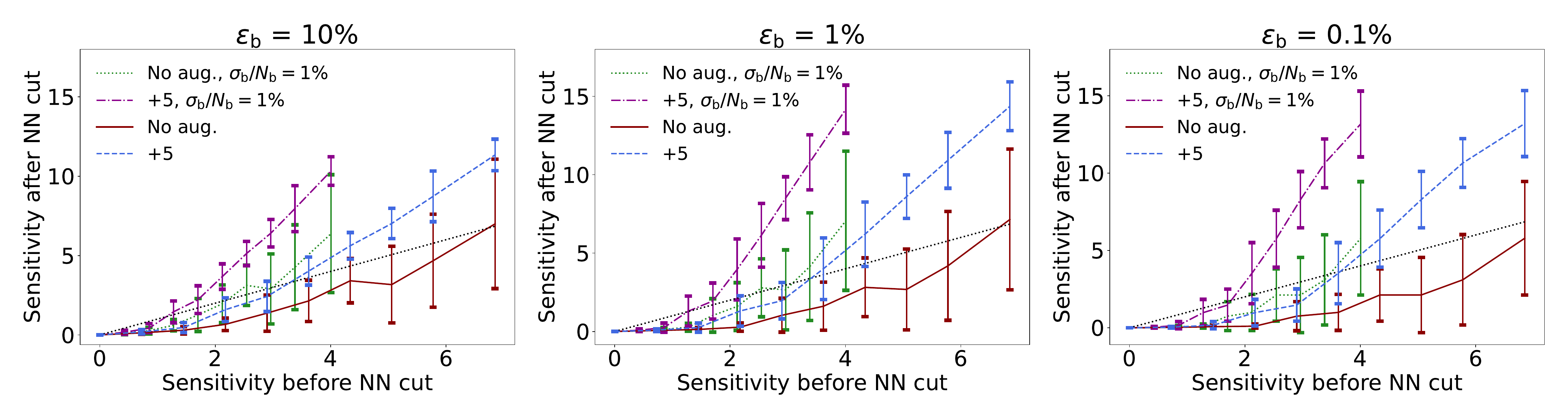}
            }
            \caption{The sensitivities before and after the NN selection when the systematic uncertainty is taken into account. The gray dotted line represents the sensitivity before NN selection. The error bar is the standard deviation of 10 times training. The augmentation method used in these plots is ``$p_{\text{T}}$ smearing + jet rotation.''}
            \label{fig:sensitivity_improvement_with_systematic_uncertainty}
        \end{figure}

        Figure~\ref{fig:sensitivity_improvement_with_systematic_uncertainty} shows the sensitivity improvement with systematic uncertainty for the ID and DD scenarios. Here, we consider a relative background uncertainty of 1\% for illustration purposes, though the typical relative uncertainty is 5\%~\cite{CMS:2019gwf}. The neural network with data augmentation still outperforms the one without data augmentation even when the systematic uncertainty is present.

        In the presence of relative uncertainty, the curves in figure~\ref{fig:sensitivity_improvement_with_systematic_uncertainty} are compressed in both the horizontal and vertical directions. However, the compression is significantly greater in the horizontal direction than in the vertical direction. This occurs because the number of background events after applying the NN cut is substantially smaller than before the NN cut. Consequently, even when systematic uncertainty is taken into account, data augmentation still significantly enhances the performance of neural networks.

    % subsection impacts_of_systematic_uncertainty (end)
% section data_augmentation (end)

\section{Conclusions}% (fold)
\label{sec:conclusions}

    Weakly supervised learning combines the benefits of both fully supervised and unsupervised approaches.  In particular, the neural networks can learn the signal properties and be trained on real data. In this work, we train a classifier on the mixed datasets constructed from the SR and SB of assumed real data. Neural networks typically require sufficiently large datasets to work properly or for better performance. This poses a challenge for collider experiments when the signal production rate is limited by luminosity. In this work, we propose using data augmentation methods to enlarge the size and diversity of the training dataset to overcome this problem.

    We utilize three physics-inspired data augmentation methods, which take into account the physical properties and the experimental resolution of the detector. By augmenting the training data with these methods, the neural networks are trained with a wider range of realistic variations and seen to gain better ability in classifying the signal and background events.

    Using the dark valley model as an explicit new physics possibility at the LHC, we show that data augmentation significantly enhances the neural network's performance, effectively reducing the learning thresholds from around $6\sigma$ to $3\sigma$ for both ID and DD scenarios defined in the main text. Moreover, the standard deviations are reduced to a half, leading to more stable and robust neural networks. The ``$p_{\text{T}}$ smearing + jet rotation'' features the best performance among the three methods.

    In summary, this study demonstrates that physical data augmentation provides an effective way to address the challenge of limited real data on which we train our neural networks in the CWoLa approach. By applying the transformations to the data based on human insights into the underlying physics, we can enlarge the training dataset by providing greater diversity, thereby significantly enhancing the neural network's ability for generalization and thus its performance. Data augmentation techniques extend beyond weakly supervised learning and can be utilized in scenarios with limited real data. This strategy unlocks new opportunities to enhance collider searches, even in the face of data scarcity.

% section conclusions (end)

\acknowledgments
We thank Hugues Beauchesne for his contributions at the early stage of this project. This work was supported by the National Science and Technology Council under Grant No. NSTC-111-2112-M-002-018-MY3.

\bibliography{biblio}
\bibliographystyle{utphys}

\end{document}